\documentclass[12pt]{article}
\usepackage{amsmath, amssymb, graphics}
\usepackage{graphicx}
\usepackage[final]{showkeys}
\usepackage{bbm}
\usepackage{color}
\usepackage{subfigure}
\usepackage{float}

\usepackage{appendix}

\usepackage{graphics, psfrag}
\newcommand{\mathsym}[1]{{}}
\def\id{\protect{{1 \kern-.28em {\rm l}}}}

\def\be{\begin{eqnarray}}
\def\ee{\end{eqnarray}}

\makeatletter
\renewcommand\section{\@startsection {section}{1}{\z@}%
                                   {-3.5ex \@plus -1ex \@minus -.2ex}%
                                   {2.3ex \@plus.2ex}%
                                   {\normalfont\large\bfseries}}
\renewcommand\subsection{\@startsection{subsection}{2}{\z@}%
                                   {-3.25ex\@plus -1ex \@minus -.2ex}%
                                   {1.5ex \@plus .2ex}%
                                   {\normalfont\normalsize\bfseries}}

\makeatother

\def\IR{\mathbb{R}}

\def\IZ{\mathbb{Z}}

\def\z{{\rho}}

\def\NeqFour{{\cal N}=4}

\def\vol{\mbox{vol}}
\def\Vol{\mbox{Vol}}
\def\Tr{{\rm Tr}}

\def\draftnote#1{{\color{red} #1}}
\def\bldraft#1{{\color{blue} #1}}

\begin{document}

\overfullrule=0pt
\parskip=2pt
\parindent=12pt
\headheight=0in \headsep=0in \topmargin=0in \oddsidemargin=0in

\vspace{ -3cm}
\thispagestyle{empty}
\vspace{-1cm}

\

\

\begin{center}
\vspace{1cm}
{\Large\bf  
A twistor string for the ABJ(M) theory
}

\end{center}

\vspace{.2cm}


\begin{center}
 
Oluf~Tang~Engelund~\footnote{ote5003@psu.edu} 
and 
Radu~Roiban~\footnote{radu@phys.psu.edu}

\end{center}

\begin{center}
{
\em 
\vskip 0.08cm
\vskip 0.08cm
Department of Physics, The Pennsylvania  State University,\\
University Park, PA 16802 , USA
}
 \end{center}


\vspace{1.5cm}

\vspace{.2cm}

\begin{abstract}

We construct an open string theory whose single-trace part of the tree-level S-matrix reproduces the S-matrix of the ABJ(M) theory
with a unitary gauge group. We also demonstrate that the multi-trace part of the string theory tree-level S-matrix -- which has no 
counterpart in the pure ${\cal N}=6$ super-Chern-Simons theory -- is due to conformal supergravity interactions and identify certain 
Lagrangian interaction terms.  
Our construction suggests that there exists a higher dimensional theory which can be dimensionally-reduced, in a certain sense, 
to the ABJ(M) theory. It also suggests a generalization of this theory to product gauge groups with more than two factors.

\end{abstract}

\newpage

\section{Introduction}

Quite generally, scattering amplitudes of quantum field theories have many different presentations  (see {\it e.g.} \cite{Elvang:2013cua} for a review) 
 and each of them exposes different properties while potentially obscuring others. 
 For example, a Feynman graph presentation of amplitudes of gauge theories
manifests locality and (at tree-level) the poles on which they factorize but obscures most symmetries (especially those that emerge 
only on shell), relations between color-ordered amplitudes, {\it etc.} 

In contrast, Witten's twistor string formulation of tree-level $\NeqFour$ super-Yang-Mills theory \cite{Witten:2003nn} 
in the connected prescription \cite{Roiban:2004yf}
makes manifest its super-conformal symmetry, the U$(1)$ decoupling identities and the Kleiss-Kuijff relations \cite{Kleiss:1988ne}, 
while making their factorization properties quite difficult to identify \cite{Vergu:2006np}. It also exposes new properties, such as the
localization on curves of a certain degree in the relevant supertwistor space \footnote{A different localization of (s)YM amplitudes, valid in 
any number of dimensions, has recently been proposed and discussed in \cite{scattering_eq}. }. 
The disconnected formulation
implies the existence of a recursive construction of amplitudes based on maximally-helicity-violating building blocks~\cite{Cachazo:2004kj}.
The Grassmaniann presentation \cite{ArkaniHamed:2009dn} of the same amplitudes manifests their dual superconformal invariance (at 
tree-level and at the level of the integrand at loop level)  \cite{Drummond:2006rz, Bern:2006ew, Drummond:2008vq} while obscuring locality.
%
Last but not least  the twistor string theory \cite{Witten:2003nn} and its open string formulation \cite{Berkovits:2004hg} suggest that there 
may exist a formulation of $\NeqFour$ conformal supergravity whose action, differently from the standard one~\cite{Bergshoeff:1980is}, 
does not have a manifest SU$(1,1)$ global symmetry~\cite{Berkovits:2004jj}.

The three-dimensional ABJ(M) theory \cite{Aharony:2008ug} shares many of the properties of $\NeqFour$ sYM theory such as the 
integrability of the planar dilatation operator \cite{ABJMintegrability} and invariance of its tree-level amplitudes and of its  loop 
integrands under the Yangian of the superconformal group \cite{Bargheer:2010hn, Huang:2010qy, Gang:2010gy}. 
It was suggested that the scattering amplitudes of this theory have a Grassmannian formulation \cite{Lee:2010du} as well as
a formulation~\cite{Huang:2012vt} exhibiting holomorphic localization on curves of a certain degree in twistor space\footnote{The analogous 
relation between the Grassmannian and twistor string formulation of ${\cal N}=4$ sYM tree-level amplitudes was discussed in \cite{ArkaniHamed:2009dg}.
The ABJ(M) amplitude expression of \cite{Huang:2012vt} was recently shown to be equivalent to an alternative integral formula which satisfies 
all factorization properties \cite{Cachazo:2013iaa}.}, 
which in turn suggests that a twistor string theory may exist for this theory as well. Such a string theory would also contain conformal 
supergravity~\footnote{The coupling of the ABJ(M) theory with ${\cal N}=6$ conformal supergravity was discussed from a field theory 
perspective in \cite{Nishimura:2013poa}. }. 
Unlike its four-dimensional counterpart, three-dimensional conformal supergravity does not have any asymptotic states~\cite{Lindstrom:1989eg} 
and therefore does not lead to factorization channels which are not present in the flat space theory. Consequently, its presence  can only be 
inferred  either from off-shell data or from 
scattering amplitudes which vanish in the absence of conformal supergravity interactions.

Generalizations of the ABJ(M) theory, to more general gauge groups \cite{Aharony:2008gk, Hosomichi:2008jb, Schnabl:2008wj} and 
to lower numbers of supercharges \cite{Gaiotto:2008sd, Hosomichi:2008jd, Schwarz:2004yj} have been discussed extensively.  
Another interesting question is whether there exist higher-dimensional theories from which the ABJ(M) theories can be obtained by 
some truncation procedure, such as dimensional reduction.

In this paper we discuss the construction of a (twistor) string theory which has the SU$(N)\times\text{SU}(M)$ ${\cal N}=6$ Chern-Simons 
theory (ABJ(M)) as a subsector to which it can be consistently truncated; the resulting scattering amplitudes take the form given in
\cite{Huang:2012vt}. \footnote{ It may also be possible to construct a string theory whose amplitudes naturally take the form \cite{Cachazo:2013iaa}
along the lines of \cite{Mason:2013sva}.} While this string theory appears to have infinitely many states, they combine into representations of 
a larger symmetry group and thus may be interpreted as describing a higher-dimensional field theory.
This construction yields the scattering amplitudes of this theory and some of its conservation laws without identifying the corresponding
Lagrangian.
%
%
It also turns out that, within our construction, it is possible to identify tree-level S-matrices of other consistent truncations of 
this putative higher-dimensional theory; they appear to correspond to ${\cal N}=6$ superconformal field theories that are not 
immediately included in the classification \cite{Schnabl:2008wj}.
We shall also discuss the consequences of conformal supergravity interactions: they generate tree-level multi-trace color-ordered 
amplitudes. For four-point amplitudes we compare explicitly the string-theory-generated expressions with those following from the 
Lagrangian  put forth in \cite{Chu:2009gi,Chu:2010fk,Nishimura:2013poa}.

The crucial difference between the twistor space of the three and four-dimensional Minkowski spaces is that, while the latter 
has a unitary symmetry group, the former has only an orthogonal one.  The orthogonality constraint may in principle be imposed at 
the level of the worldsheet action at the expense of introducing  ghost fields. Alternatively, at tree-level it may also be imposed only as a choice of 
the kinematics determining the scattering states. More precisely, the string theory we shall construct has SU$(3,2|4,1)$ 
symmetry and restricting to a particular subset of states with vanishing charge under a certain worldsheet global U$(1)$ symmetry 
yields the tree-level scattering amplitudes of the ABJ(M) theory. \footnote{While this truncation is natural and consistent at tree level, loop 
amplitudes of chargeless states receive contributions from charged states as well. We shall not discuss loop amplitudes in this paper.}
For general states, the amplitudes of this theory receive contributions from worldsheet instantons of all degrees;
upon restricting to states transforming only under the expected $\text{SU}(3)$ on-shell R-symmetry of the ABJ(M) theory the 
only contributing instantons are those of degree equal to half the number of vertex operators, as suggested by \cite{Huang:2012vt}.
The localization on instantons of fixed degree is realized through holomorphic delta functions  \cite{Witten:2004cp, Roiban:2004yf} 
\footnote{\label{defholodelta}
See  \cite {Cachazo:2005ga} for a detailed discussion on holomorphic delta functions.
%
},
in close similarity with the twistor string for the $\NeqFour$ sYM theory.  

As in all theories with manifest symmetries, restriction to a subset of states which are invariant under a subgroup of the symmetry group 
leads to singular or vanishing factors in scattering amplitudes, depending on whether the inert generators are Grassmann-even or odd, 
respectively. For example, in a four-dimensional field theory, restricting to three-dimensional kinematics specified by {\it e.g.} the vanishing 
of the third component of the momentum yields a factor of $\delta(0)\equiv\delta(\sum_{i=1}^nP_i^3)$. Similarly, if the states are invariant 
under some on-shell supersymmetry generator $Q = m^\alpha_A q^A_\alpha$ with some fixed matrix $m$, one finds a factor of 
$0=\delta(0)=\delta(\sum_{i=1}^n m^\alpha_A (q_i)^A_\alpha ))$.
Thus, to extract the scattering amplitudes of such invariant states it is necessary to identify the generators leaving the states invariant 
 and extract the corresponding vanishing or singular factors. 

This paper is organized as follows. In \S~\ref{embedding} we discuss an embedding of the twistor space of the three-dimensional 
Minkowski space into (a noncompact form of) CP$^{4|5}$. In \S~\ref{string_th} we construct an open string theory on CP$^{4|5}$ and its 
states and discuss the truncation of its space of states to the ABJ(M) spectrum as well as the general structure of scattering amplitudes 
of this theory, following closely \cite{Berkovits:2004hg}. In \S~\ref{ABJMamplitudes} we construct the scattering amplitudes of the 
constrained states and recover the expression of \cite{Huang:2012vt}.  In \S~\ref{CSG} we discuss the contribution of
conformal supergravity states to amplitudes and illustrate it by computing the four-point amplitudes and comparing them with the 
ones following from the Lagrangian proposed in \cite{Nishimura:2013poa}. We also discuss a similar comparison for certain 
arbitrary-multiplicity amplitudes.
We close in \S~\ref{outlook} with remarks on various extensions of our construction to gauge groups with more than two factors,
the inclusion of other states and the higher-dimensional interpretation of our construction as well as on the possibility of enforcing the 
truncation to the ABJ(M) spectrum at the quantum level.

\section{An embedding of the 3d twistor space  \label{embedding}}


The supertwistors for the ${\cal N}$-extended three-dimensional superconformal group OSp$({\cal N}|4,\IR)$ were discussed in detail 
in \cite{Huang:2010rn}; they are given by the pairs $(\xi^\mu,\eta^A)$ where the two components transform in the fundamental 
representations of Sp$(4)$ and SO$({\cal N})$, respectively. They are real and self-conjugate, 
%
%
\be
[\xi^\mu,\xi^\nu] = \Omega^{\mu\nu}
\quad, \qquad
\{\eta^A,\eta^B\} = \delta^{AB} \ ;
\label{TWconstraint}
\ee 
From this perspective the twistor space may be interpreted as a phase space on which one may  
choose Darboux-like coordinates, $\xi^\mu=(\lambda_a,\mu_b)$ such that
\be
[\lambda_a,\mu_b]=\epsilon_{ab} \ .
\label{darboux}
\ee
Thus, (wave) functions on this space depend only on half of the coordinates, {\it e.g.} $\lambda^a$. The generators of OSp$({\cal N}|4)$  
are second-order differential operators \cite{Huang:2010rn}. Unlike the twistor space of the four-dimensional Minkowski space, the 
properties of spinors make it difficult to linearize the action of the (super)conformal group. \footnote{{\it E.g.} momenta are still bilinears 
in $\lambda$: $P^{\alpha\beta}=\lambda^\alpha\lambda^\beta$.}

To have more manifest symmetry we may relax the constraint \eqref{TWconstraint} embed this space as a hypersurface in a larger one; 
we subsequently impose this constraint on (wave) functions defined on this larger space. Let us begin by discussing the bosonic 
coordinates and append afterwards the Grassmann directions. 

We interpret the bosonic coordinates $\xi^\mu=(\lambda_a,\mu_b)$ as part of the coordinates of CP$^{2, 2}$ \footnote{One may in 
principle consider larger spaces, such as CP$^{2+n, 2}$ with $n>0$. It is however not clear whether this will make further considerations 
more natural.}.  The homogeneous coordinates, 
\be
y_I=(\lambda_\alpha,\mu_{\dot a}) ~~,\quad \alpha=1,2,3\equiv (a, 3) ~,~~{\dot a}  =1,2 \ ,  
\ee
transform in the fundamental representation of SU$(3,2)$. The three-dimensional conformal group, embedded as
Sp$(4,\IR)\subset \text{SU}(2,2)\subset \text{SU}(3,2)$, acts linearly on the first two and the last two 
components of $y_I$. We denote by $z^I$ the canonical conjugates of $y_I$,  
\be
{\tilde f}(z^I) = \int d^{4+n} y \, e^{iz^I y_I}\, f(y_I)~~, \quad z^I = ({\bar \mu}^\alpha, {\bar\lambda}^{\dot a})\ .
\ee
In the phase space $(y,z)$, the regular twistor space is obtained as the hypersurface identifying $\mu$ and ${\bar\mu}$, {\it i.e.} as
(a representative of) the solution to the constraint
\be
\epsilon^{{\dot a} a}\mu_{\dot a} {\bar\mu}_a  = 0 \ .
\label{external_state_condition}
\ee
located at a point in ${\bar \mu}^3$. This identification, which breaks scale invariance of the complex projective space to only a $\IZ_2$, 
simply enforces eq.~\eqref{darboux} which implies that $\lambda$ and $\mu$ are canonically conjugate to each other. 

{}From a bosonic point of view we shall construct a string theory with target space $(z, y)$ and impose the identification as a specific choice 
of asymptotic state kinematics. While this string theory has the complete SU$(3,2)$ symmetry, vertex operators for the restricted states do 
not as the additional ${\bar \mu}^3$ direction is treated separately. 

It is not difficult to include the Grassmann coordinates in this construction; we shall focus on the case ${\cal N}=6$, for which the 
two types of supertwistor space wave functions are
\be
{\hat \Phi}(\lambda,\eta)&=&\phi(\lambda) +\psi_I(\lambda)\eta^I+\phi_{IJ}(\lambda)\eta^I\eta^J+\psi_{IJK}(\lambda)\eta^I\eta^J\eta^K \ ,
\nonumber\\
{\hat \Psi}^{IJK}(\lambda,\eta)&=&{\bar \psi}^{IJK}(\lambda) +\eta^I{\bar \phi}^{JK}(\lambda)+\eta^I\eta^J{\bar\psi}^K(\lambda)
+\eta^I\eta^J\eta^K{\bar\phi}(\lambda) \ ,
\label{superfields}
\ee
with $I,J,K=1,2,3$. We will denote the fermionic completion of $y$ by $\eta$ and that of 
$z$ by $\bar\eta$; similarly to $z$ and $y$, they are Fourier-conjugates of each other.  
To this end we should embed SO$(6)$ into a larger 
symmetry group and partially break it while preserving the manifest SU$(3)$ on-shell symmetry (of asymptotic states) as well as the 
non-manifest SO$(6)$ symmetry (of amplitudes). 
%
The suitable choice is coupled to the way 
it is broken and the wave functions on the twistor space are recovered.
Since the bosonic part of the supertwistor space is obtained through an SL$(2,\IR)$-invariant constraint \eqref{external_state_condition}
we may consider a similar constraint involving the Grassmann-odd coordinates, such as the vanishing of a supermomentum component
similar to  \eqref{external_state_condition}.  It turns out that for our purpose it is useful to choose 
\be
SU(4)\subset SU(4,1) \ ,
\label{larger_group}
\ee
such that the complete symmetry of the embedding space is SU$(3,2|4,1)$. Then, we break the SU$(4,1)$ to its SU$(3)$ subgroup 
by identifying SU$(1,1)\subset \text{SU}(4,1)$ with an SU$(1,1)\subset\text{Sp}(4,\IR)$ and imposing the vanishing of an SU$(1,1)$-invariant 
supercharge,  
\begin{align}
{\mu}_{\dot 1}{\bar\eta}^4+{\mu}_{\dot 2}{\bar\eta}^5 = 0 \ .
\label{zero_super_momentum}
\end{align}
It is in principle possible to choose a symmetry group larger than in eq.~\eqref{larger_group}; however, constraints breaking it to the requisite 
SU$(3)$ R-symmetry of the three-dimensional ${\cal N}=6$ on-shell superspace appear to be less natural.

To summarize, we shall construct a string theory with target space CP$^{2,2|4,1}$ and choose to express the kinematic information of the vertex operators 
in terms of the coordinates $({\bar\mu},\mu, {\bar \eta})$. We then obtain the states of a theory with OSp$(6|4)$
symmetry by imposing the following constraints on the external state kinematics:

\begin{enumerate}

\item We fix ${\bar \mu}^3$ to some value and impose eq.~\eqref{external_state_condition} in the form 
\be
\mu_{\dot{1}}=\bar{\mu}_1
~~, \qquad 
\mu_{\dot{2}}=\bar{\mu}_2 \ .
\label{external_state_condition_av}
\ee
As discussed in the Introduction, such a kinematic configuration is quite singular. Ignoring fermions, it is akin to setting to zero one 
component of all external momenta. In the case at hand we expect the singular factor  
\be
\delta(0)\equiv \delta(\sum_i \epsilon^{{\dot a} a} \mu_{i{\dot a}}{\bar\mu}_{i a}) \ ,
\label{expect_singular}
\ee 
where the sum runs over all external states.  We will be interested in the coefficient of  such a singular factor.

\item Impose the vanishing of the supercharge component \eqref{zero_super_momentum}. 
We will do this by applying to the result of item 1) the projector
\be
{\cal P}_F(\bullet)=\int d {\bar \eta}^4d{\bar \eta}^5 \delta({\mu}_{\dot 1}{\bar\eta}^4+{\mu}_{\dot 2}{\bar\eta}^5)(\bullet) \ .
\label{projector}
\ee
Given a function $G$,  ${\cal P}_F(G)$ is a particular combination of 
$\partial_{{\bar\eta}^4}G|_{{\bar\eta}^{4,5}=0}$ and  $\partial_{{\bar\eta}^5}G|_{{\bar\eta}^{4,5}=0}$. 
The functions produced by this projector are charged under the U$(1)$ 
that rephases uniformly ${\bar\eta}^{4}$ and ${\bar\eta}^{5}$.
While this projector preserves an SU$(1,1)$ symmetry, explicitly solving the constraint imposed by it will break it; this 
is a reflection of the fact that only one SL$(2,\IR)\subset \text{Sp}(4,\IR)$ acts manifestly on functions on the 
three-dimensional twistor space. The integration over ${\bar\eta}^{4,5}$ restores this symmetry, albeit non-manifestly.

Similarly to the bosonic constraint, we should expect that imposing the vanishing of a component of the supercharge is a rather 
singular limit and that we should find a Grassmann delta function $\delta(0)$ whose argument is in fact the corresponding component 
of the supercharge; such a delta function vanishes due to its Grassmann 
nature. Unlike the bosonic case however we impose the vanishing of a supercharge components through the integral operator 
\eqref{projector} which eliminates the constrained Grassmann variables; we should therefore expect a different manifestation -- though 
similar in spirit -- of the vanishing Grassmann delta function. The only possibility that is independent of the number of vertex operators is
\be
\delta(0) \rightarrow \sum_i \epsilon^{{\dot a} a} \mu_{i{\dot a}}{\bar\mu}_{i a} \ .
\label{vanishing_factor}
\ee
We shall see that this expectation is indeed realized.

\end{enumerate}



\section{A string theory on CP$^{2,2|4,1}$ and its truncation to ABJ(M) states  \label{string_th}}

With the ingredients discussed in the previous section we now construct, following \cite{Berkovits:2004hg}, an open string 
theory\footnote{Presumably, a heterotic theory can also be constructed along the lines of \cite{Mason:2007zv}.} 
on CP$^{2,2|4,1}$ and its vertex operators that correspond to the ABJ(M) states. 
The worldsheet fields and their conjugates, \footnote{We shall use the same notation for the worldsheet fields as for external state kinematics.
To avoid potential confusion, we shall include explicitly the worldsheet position as argument of the worldsheet fields.}
\be
Z^I(\z)=(\bar{\mu}^\alpha(\z),\bar{\lambda}^{\dot{a}}(\z),\bar{\psi}^A(\z))\quad ,\qquad Y_I(\z)=(\lambda_\alpha(\z),\mu_{\dot{a}}(\z),\psi_A(\z)) \ ,
\label{freefields}
\ee
transform in the fundamental representation of SU$(3,2|4,1)$ ($\alpha=1,2,3, {\dot a}=1,2$, $A=1,\dots,5$.). Their dimensions are $(0,1)$ respectively
and have a standard first-order action
\begin{align}
S&=\int d^2 \z \, (Y_{LI}\nabla_R Z^I_L+Y_{RI}\nabla_L Z^I_R)+S_G \ .
\label{YZaction}
\end{align}
Similarly to \cite{Berkovits:2004hg}, the covariant derivative $\nabla$ contains the connection for the local GL$(1)$ symmetry acting as 
$Z\rightarrow t Z$, $Y\rightarrow t^{-1} Y$ which relates the free fields \eqref{freefields} and the coordinates  on CP$^{2,2|4,1}$.

The term $S_G$ in \eqref{YZaction} is the action for a collection of two-dimensional fermions that will be responsible for 
the target space gauge degrees of freedom. They consist of $N$ dimension-$1/2$ fields $\Psi_1$ and 
$M$ dimension-$1/2$ fields $\Psi_2$ with a standard action
\be
S_G = \int d^2 \z \, \sum_{i=1}^N ({\bar\Psi}^i_{1, L} \partial_R {\Psi}_{1,i L} +
{\bar\Psi}^i_{1, R} \partial_L {\Psi}_{1, i R}   + \sum_{j=1}^M {\bar\Psi}^j_{2, L} \partial_R {\Psi}_{2, j  L} +{\bar\Psi}^j_{2, R} \partial_L {\Psi}_{2,j R} ) \ .
\label{Gaction}
\ee

Quantization of \eqref{YZaction} yields the usual diffeomorphism ghosts $b,c$ and GL$(1)$ ghosts $u,v$; the corresponding 
left-moving stress tensor and GL$(1)$ current  are
\be
T_0 = Y_{LI}\nabla_L Z^I_L  + T_G + b_L\partial c_L +\partial_L(b_Lc_L) + u_L\partial v_L  
\quad, \qquad
J_{\text{GL}(1)} = Y_{LI} Z^I_L  \ .
\label{TandJ}
 \ee
The equal number of bosonic and fermionic directions in the fundamental representation of SU$(2,3|4,1)$ guarantees 
that this current is non-anomalous and  thus can be gauged. The BRST operator takes the usual form:
\be
Q = \int d\z\;\big[cT +c T_{uv} + v J_{\text{GL}(1)}  + cb\partial_L c  \,\big]
\ee
The $(Y,Z)$ system has vanishing central charge; the central charge of the ghost systems is $-26-2=-28$ and should be cancelled 
by the fermion system. Similarly to the discussion in \cite{Berkovits:2004hg, Berkovits:2004jj}, this restricts the possible gauge symmetry
of the target space effective theory. Classically however this cancellation is unimportant and we can pick any desired numbers $N$/$M$ of fermions.

Apart from the symmetries acting on $Y$ and $Z$, the action \eqref{YZaction} and \eqref{Gaction} is also invariant under unitary 
transformations acting independently on $\Psi_1$ and $\Psi_2$. As usual, the traceless part of the corresponding left-moving currents have no 
anomalous term in their OPE with the stress tensor and thus imply that correlation functions are $\text{SU}(N)\times\text{SU}(M)$-invariant.
Of the remaining U$(1)$ currents, their difference
\be
J_F = q_{} \big( \frac{1}{N}{\bar\Psi}_1\Psi_1 - \frac{1}{M}{\bar\Psi}_2\Psi_2 \big)
\ee
also has vanishing mixed anomaly and thus require that nonvanishing correlation functions have vanishing charge.

On this theory we impose open-string boundary conditions, 
\be
Y_L=Y_R
~,\quad
Z_L=Z_R
~,\quad
\Psi_{i,L}=\Psi_{i,R} \ ,
\ee
and thus vertex operators depend on a single set of variables which we call $Z=Z_L=Z_R$.

Before choosing external states to lie in the three-dimensional twistor space (and thus choosing to treat the pair 
$(\lambda_3(\z), \mu_{\dot 3}(\z))$ differently) the theory -- and its amplitudes -- are invariant under SU$(3,2|4,1)$. 
The $\Psi_i$-independent part of vertex operators have the same structure as in \cite{Berkovits:2004hg, Berkovits:2004tx}, 
which is determined by $Q$-closure and dimension constraints. The ones to be dressed with various combinations of $\Psi$ are
\footnote{The numerical matrices $T^a$ generate SU$(N+M)$ and the $\Psi$-dependent currents written here generate the 
same algebra; for later purposes we write it here in an $\text{SU}(N)\times \text{SU}(M)$ decomposition.}
\be
U&=&\int \frac{d\xi}{\xi} \delta^3(\bar{\mu}-\xi\bar{\mu}(\z))e^{i\mu_{i\dot{a}}\xi\bar{\lambda}^{\dot{a}}(\z)}
\delta^{0|5}({\bar\eta}^A - \xi\bar{\psi}^A(\z))  \ ,
\\
V_{ij}(\z)&=& U(\z)J_{ij}
~,\quad
J_{ij}^a={\bar\Psi}_iT^a\Psi_j
\ ,
\label{BigSym}
\ee
while those that do not require such dressing and thus describe a sector of gauge-singlet states~\footnote{In theories for 
which a target space Lagrangian is known such states describe conformal supergravity; in the following we shall generically 
refer to them as "conformal supergravity-like states" even though a target space Lagrangian may not be known.} are
\footnote{One can eliminate $\xi$  and express $U$ in terms of ratios ${\bar\mu_2}(\z)/{\bar\mu_1}(\z)$, {\it etc}. \cite{Berkovits:2004hg}. }
\begin{equation}
V_f = f^I(Z(\z))Y_I(\z)~~~~~~~~V_g = g_I(Z(\z))\partial Z^I(\z) \ .
\label{CSGvop}
\end{equation}
$V_f$ and $V_g$ are primary operators if the functions $f^I$ and $g_I$ solve the constraints
\be
Z^Ig_I=0\qquad \partial_If^I=0 \ ,
\ee
as discussed in \cite{Witten:2004cp, Berkovits:2004jj}.

The scattering amplitudes of this string theory are given by the standard construction,
\be
{\cal A}_n(1\dots n) = \langle cV_1(\z_1)cV_2(\z_2)cV_3(\z_3)\int d\z_4 V_4(\z_4)\dots \int d\z_n V_n(\z_n) \rangle 
\ ;
\label{general_amplitude}
\ee
as in the case of the similar open string theory describing the tree-level $\NeqFour$ sYM theory \cite{Berkovits:2004hg}, they break up into 
a sum over sectors labeled by the GL$(1)$ instanton number. While the complete tree-level S-matrix of this theory can easily be written down, it 
is not clear what quantum field theory it corresponds to. It would be interesting to analyze their structure and understand whether or not it 
is a standard field theory \footnote{The strange signature of the target space can be undone through appropriate Wick rotations.}.

\subsection{Vertex operators with reduced symmetry}

Treating differently the target space kinematics corresponding to the pair $(\lambda_3, {\bar\mu}^{3})$ and $(\psi_4, \psi_5)$ breaks the 
SU$(3,2|4,1)$ symmetry (to SU$(2,2|3)$) and introduces a few more possibilities for vertex operators, akin to vertex operators for 
Kaluza-Klein states. It also implies that, apart from the GL$(1)$ current \eqref{TandJ} it makes sense to consider a current constructed 
from these fields, 
\be
J'(\z)=q_B {\bar\mu}^3\lambda_3(\z) +q_F \sum_{A=4}^5 {\bar\psi}^A\psi_A(\z)  \ .
\label{extraJ}
\ee
Its mixed anomaly vanishes if 
\be
q_B=2q_F \ ;
\label{qBqF}
\ee
for such a choice of $q_B$ and $q_F$ non-vanishing correlation functions have vanishing charge $q'$ with respect to the current $J'$. 

Breaking SU$(3,2)$ to SU$(2,2)$ allows us to construct a family of dimension-zero operators 
\be
U_n({\bar\mu}_{a}, \mu_{\dot a},{\bar\eta};\z) = \int \frac{d\xi}{\xi}(\xi {\bar\mu}^3(\z))^n
\delta^2(\bar{\mu}-\xi \bar{\mu}(\z))e^{i\mu_{\dot{a}}\xi \bar{\lambda}^{\dot{a}}(\z)}
\delta^{0|5}({\bar\eta}^A - \xi \bar{\psi}^A(\z))  
\ ,
\label{Uvop}
\ee
which  are to be dressed with the $\Psi$-dependent dimension-one currents $J_{ij}^a$ defined in eq.~\eqref{BigSym}. These
operators may be interpreted as being the Kaluza-Klein modes of the operator $U(\z)$ in eq.~\eqref{BigSym}. The external 
kinematics is specified by ${\bar\mu}_{a}$,  $\mu_{\dot a}$ and ${\bar\eta}$. 

A further vertex operator may be obtained by localizing \eqref{BigSym} on the point ${\bar\mu}^3=0$:
\be
U_\delta({\bar\mu}_{a}, \mu_{\dot a},{\bar\eta};\z) = \int \frac{d\xi}{\xi} 
\delta^2(\bar{\mu} - \bar{\mu}(\z))\delta(\xi {\bar\mu}^3(\z))e^{i\mu_{\dot{a}}\xi \bar{\lambda}^{\dot{a}}(\z)}
\delta^{0|5}({\bar\eta}^A - \xi \bar{\psi}^A(\z)) \ .
\label{Udelta}
\nonumber
\ee
This operator carries charge $q'=-1$ with respect to the current in eq.~\eqref{extraJ} and may be interpreted {\it e.g.} as a (singular) 
limit of a superposition of the operators in eq.~\eqref{Uvop}, using a limit representation for the Dirac $\delta$ function. Thus, we may trade
$U_{-1}$ for this operator.
It may also be possible to interpret it as a twisted sector vertex operator for a $\IZ_2$ orbifold acting as $\IZ_2:{\bar\mu}^3 \mapsto - {\bar\mu}^3$;
such an orbifold would project out  $U_{-1}$. While potentially interesting, we shall not pursue this interpretation here.

In addition, there are vertex operators which do not depend on $\Psi_i$ and are analogous to those in eq.~\eqref{CSGvop} apart for the 
dependence on ${\bar\mu}^3$ which is taken as above.

Since the SU$(4,1)$ symmetry is unbroken at this stage, the target 
space Grassmann coordinates $\eta$ enter only in the combination $\delta^{0|5}({\bar\eta}^A - \xi \bar{\psi}^A(\z))$. To project the fermionic kinematics onto the desired one corresponding to states depending on three Grassmann variables 
we act with the projector ${\cal P}_F$ introduced in eq.~\eqref{projector}; for all of the operators above their fermionic part becomes
\be
u({\bar\eta},\xi{\bar\psi})={\cal P}_F \delta^{0|5}({\bar\eta}^A - \xi \bar{\psi}^A(\z)) = 
\delta^{0|3}({\bar\eta}^A - \xi\bar{\psi}^A(\z)) \delta(\xi \bar{\psi}^4(\z){\mu}_{\dot 1}+\xi\bar{\psi}^5(\z){\mu}_{\dot 2}) \ .
\label{FermiPart}
\ee
We notice that it carries nontrivial charge under the current $J'$ defined in eq.~\eqref{extraJ} and thus, since after projection 
all vertex operators depend on $\eta^A$ through $u({\bar\eta},\xi{\bar\psi})$, they are all charged under~$J'$.

To understand some of the consequences of the conservation of this current we list the charges $q'$ 
of the vertex operators thus constructed:
\be
\nonumber
q' (u({\bar\eta},\xi{\bar\psi})) &=&  q_F 
\qquad\qquad\quad~~
q'({\cal P}_FV_g) = q'({\cal P}_FV_f) = q_F
\\
q'({\cal P}_FU_n) = nq_B + q_F &=& (2n+1)q_F
\qquad
q'({\cal P}_FU_\delta) = -q_B + q_F = -q_F \ .
\label{q'charges}
\ee 
An interesting class of correlators are those containing an equal number of ${\cal P}_FU_{-1}$ and ${\cal P}_FU_0$; it turns out 
that it is possible to consistently truncate the theory to only the states described by them. We shall revisit this truncation 
in \S~\ref{outlook}; to justify it we examine the new current 
\be
J_{\text{U}(1)} = J'+J_F \ .
\label{U1_current}
\ee
For a particular choice of $q_{}$ charge for the fermions $\Psi_i$,
\be
q_{}=-\frac{NM}{N+M}q_F \ ,
\label{fermion_charge}
\ee
the appropriately-dressed operators ${\cal P}_FU_{-1}$ and ${\cal P}_FU_0$,
\begin{align}
{V}_{-1, 12}({\bar\mu}_{a}, \mu_{\dot a},{\bar\eta};\z)=& J_{12} \; {\cal P}_FU_{-1}({\bar\mu}_{a}, \mu_{\dot a},{\bar\eta};\z)
\qquad
{V}_{0,21}({\bar\mu}_{a}, \mu_{\dot a},{\bar\eta};\z)=  J_{21} \; {\cal P}_FU_0({\bar\mu}_{a}, \mu_{\dot a},{\bar\eta};\z) \ ,
\label{ABJMvop}
\end{align}
have vanishing charge with respect to it. All other vertex operators  that can be constructed have nonzero $q_{\text{U}(1)}$ charge.
Therefore,  from the perspective of the target space effective action for this theory it is possible to consistently truncate it {\it at tree 
level} to the states \eqref{ABJMvop}. Indeed, introducing charged operators in a correlation function of chargeless operators requires in 
fact the introduction of two operators of opposite charges and thus no chargeless source can appear in the equations of motion for 
charged fields; thus, the former can be consistently truncated away.
We shall comment in \S~\ref{outlook} on the possibility of truncating to chargeless states at higher orders in perturbation theory.

We also note that the chargeless states described by the vertex operators \eqref{ABJMvop} can be put in one to one correspondence 
with the states of the two ABJ(M) multiplets \eqref{superfields}. The absence of vertex operators that are independent of $\Psi_i$ is 
consistent with the expectation that the only asymptotic states of this theory correspond to fields transforming nontrivially under the 
gauge group. 
To construct these operators we have chosen the perhaps unorthodox choice of labeling the external kinematics with half the 
coordinates and with the momenta conjugate to the other half of the coordinates. This choice makes it straightforward to impose on 
these operators the constraint in eq.~\eqref{external_state_condition_av}, enforcing the fact that the twistor space 
is a phase space. We shall not do this here, but rather include it in their correlation functions.

\subsection{The general structure of scattering amplitudes}

Color-dressed amplitudes are computed in the usual way, from disk correlation functions of integrated and unintegrated vertex operators. As in 
the case of the $\NeqFour$ open string as well as in the case of amplitudes \eqref{BigSym} with SU$(3,2|4,1)$ symmetry, 
due to the presence of the GL$(1)$ gauge field, correlation functions are given by sums over sectors labeled by the instanton number 
\cite{Berkovits:2004hg, Berkovits:2004tx},
\be
{\cal A}_{2n}(1\dots 2n)\!\! &=&\!\! \langle cV_{-1}(\z_1)cV_0(\z_2)cV_{-1}(\z_3)\int d\z_4 V_0(\z_4)\dots \int d\z_{2n} V_0(\z_{2n}) \rangle 
=\sum_{k=1}^\infty {\cal A}_{n,k}
\\
{\cal A}_{2n,k}\!\!&=&\!\! \langle cV_{-1}(\z_1)cV_0(\z_2)cV_{-1}(\z_3)\int d\z_4 V_0(\z_4)\dots \int d\z_{2n} V_0(\z_{2n}) \rangle_k  \ .
\label{amplitudes_general}
\ee
The currents $J_{12}$ and $J_{21}$ in the definition of $V_{-1}$ and $V_0$ in eq.~\eqref{ABJMvop} introduce the alternate ordering 
used here.
The integration over the auxiliary $\xi$ variable in each vertex operator effectively imposes a gauge-fixing 
of GL$(1)$. In the sector with instanton number $k-1$,  $Z$ has $k$ zero modes; since there is no non-trivial 
OPE between the $U_0$ and $U_{-1}$ factors, their correlation function is solely given by an integral over 
these zero modes. The expression of $Z$ at the position of each vertex operators is,  
\be
\bar{\mu}^\alpha(\z_i) =\sum_{m=1}^k z_{(m)}^\alpha              \z_i^{m-1}
~~,\quad
\bar{\lambda}^{\dot{a}}(\z_i)=\sum_{m=1}^k   z_{(m)}^{\dot{a}}  \z_i^{m-1}
~~,\quad
\bar{\psi}^A(\z_i)  =\sum_{m=1}^k z_{(m)}^A              \z_i^{m-1}
\ ;
\label{instanton_psi}
\ee
and the integration measure over the $k$ collective coordinates $(z_{(m)}^\alpha,z_{(m)}^{\dot a},z_{(m)}^A)$ is
\be
\int \prod_{m=1}^k dz_{(m)}^{5|5}\;\bullet \ .
\ee
To completely define the integral it is necessary to also specify the integration contour;
we will take it such that it instructs us to sum over the zeroes of the arguments of the delta functions and 
requires use of Cauchy's theorem for the other singularities of the integrand. 
A similar contour
is chosen if $V_{\delta}$ is used instead of $V_{-1}$.

A common part to all correlation functions determining scattering amplitudes of fields charged under the gauge group is
the correlator of currents $J_{ij}$ \eqref{BigSym}. They determine the various trace structures of amplitudes. 
We will be interested mainly in the vertex operators \eqref{ABJMvop}, so we sketch here the structure of the correlator of $n$ 
$J_{12}$ and $n$ $J_{21}$ currents. Since $\langle \Psi_i\Psi_j\rangle\propto \delta_{ij}$, the insertion points of the operators 
$J_{12}$ and $J_{21}$ alternates on the boundary of the disk. Thus,
\be
\label{current_correlator}
\langle J^{a_1}_{12}(\z_1)\dots J^{a_{2n}}_{21}(\z_{2n}) \rangle \!\!&=&\!\! \sum_{{\cal S}_1}
\frac{\Tr[T^{a_1}T^{a_2}\dots T^{a_{2n}}]}{
{\rm cyc}(1,\dots,2n)
}
+\sum_{l=1}^{n-1}  \sum_{{\cal S}_2^l}
\frac{\Tr[T^{a_1} \dots T^{a_{2l}}]\Tr[T^{a_{2l+1}}\dots T^{a_{2n}}] }{
{\rm cyc}(1,\dots,2l){\rm cyc}(2l+1,\dots,2n)
}
  \nonumber
 \\
 &&\qquad\qquad
 +\text{triple-traces}+\dots \ ,
\nonumber
 \\
 { \rm cyc}(\z_l,\dots,\z_m)&=&(\z_m-\z_l)\prod_{i=l}^{m-1}(\z_i-\z_{i+1})
\ee 
where the sums run over the permutations of even and odd labels up to cyclicity:
\be
{\cal S}_1 &=& \big(S(1,3,\dots,2n-1)\times S(2,4,\dots,2n)\big)/\IZ_{2n}
\\
{\cal S}^l_2 &=& \big(S(1,3,\dots,2n-1)\times S(2,4,\dots,2n)\big)/\IZ_{2l}\times \IZ_{2(n-l)}
\quad
etc.
\nonumber
\ee
Inclusion of $J_{11}$ and $J_{22}$ factors is trivial, albeit somewhat cumbersome to write out in general. 
While from a field theory perspective amplitudes with a single-trace color structure are expected to receive contributions only 
from fields transforming nontrivially under the gauge group, the amplitudes with a multi-trace  structure receive contributions
from the exchange of gauge-singlets,
such as perhaps conformal supergravity. 
In the case of the ABJ(M) theory it was argued \cite{Bargheer:2010hn} that the entire tree-level S matrix has only the former color 
structure; thus, all contribution to the latter structures necessarily comes only from such singlet states.

\section{ABJ(M)  amplitudes from the truncated CP$^{2,2|4,1}$ string theory \label{ABJMamplitudes} }


Let us now restrict ourselves to the vertex operators \eqref{ABJMvop} and evaluate the integrals over moduli 
of the fields ${\bar\psi}^4(\z)$, ${\bar\psi}^5(\z)$, ${\bar\mu}^3(\z)$ and ${\bar\lambda}^{\dot a}(\z)$ which correspond to the 
directions of CP$^{2,2|4,1}$ that are orthogonal to the hypersurface representing the three-dimensional twistor 
space. 

The integral over the moduli  of ${\bar\psi}^{4,5}$ constrains the degree $k$ of the  instanton and relates it to the number of vertex operators.
Indeed, the only part of the vertex operators containing these moduli is the result \eqref{FermiPart} of the projection operator. 
For a fixed number $2n$ of vertex operators there are $2n$ such delta functions. Since they are linear in the $2\times k$ fermionic moduli
 ($k$ for each one of the two fermionic directions),  the only nonzero contribution to their correlation function comes from instantons with degree 
\be
k={n} \ .
\label{instanton_degree}
\ee
Thus,  the $2n$-point amplitude is
\be
{\cal A}_{2n, k}&=& \delta_{k, n} {\cal A}_{2n,n}\quad \text{with}
\cr
{\cal A}_{2n, n} &=& \int \frac{\prod_{i=1}^{2n}d\z_i}{\Vol\,\text{GL}(2)} \prod_{m=1}^{k} d^{5|5}dz_{(m)}^I
\prod_{i\in \{V_{-1}\}}U_{-1}(\z_i)\prod_{j\in \{V_0\}}U_0(\z_j)\; \langle J_{12}(\z_1)\dots J_{21}(\z_{2n}) \rangle~,~~
\label{theamplitude}
\ee
where $\{V_{-1}\}\equiv\{1, 3, 5,\dots \}$ and $\{V_0\}\equiv\{2, 4, 6,\dots \}$ stand for the set of labels of operators of type $V_{-1}$ and $V_0$, respectively,
and $\langle J_{12}(\z_1)\dots J_{21}(\z_{2n}) \rangle$ is given by \eqref{current_correlator}. In the following we continue to denote the 
instanton degree by $k$.

Next, we impose the kinematic restrictions and evaluate the integrals over the variables that do not appear in the three-dimensional twistor space.
To this end we parametrize ${\cal A}_{2n, n}$ as
\be
\label{Aparametrization}
{\cal A}_{2n, n} = \int \frac{\prod_{i=1}^nd\z_i}{\Vol\,\text{GL}(2)} \frac{d\xi_i}{\xi_i} \prod_{m=1}^{k} d^{2|3}dz_{(m)}^I\; I_3I_\text{exp}I_{4,5}
\prod_{i=1}^{2n}\delta^{0|3}({\bar\eta}_i^A - \xi_i \bar{\psi}^A(\z_i))
\; \langle J_{12}(\z_1)\dots J_{21}(\z_{2n}) \rangle
\ee
where $I_3$ contains the integral over  the ${\bar\mu}^3(\z)$ moduli, $I_\text{exp}$ contains the integral over the ${\bar\lambda}$ moduli 
in the presence of the kinematic constraint \eqref{external_state_condition_av} (that half of the three-dimensional twistor space directions 
are to be interpreted as the conjugates of the other half) while $I_{45}$ is the result of the integration over the 
${\bar \psi}^4$ and ${\bar \psi}^5$ moduli in the presence of the kinematic constraint \eqref{external_state_condition_av}  as well as 
of the constraints generated in the evaluation of $I_\text{exp}$. We shall evaluate the various factors in this order.

The integral $I_3$ over $z_{(m)}^3$ moduli parametrizing  ${\bar\mu}^3(\z)$ is trivial because the only dependence on these variables
is due to the $(\xi_i{\bar\mu}^3(\z_i))^{-1}$ factors in $V_{-1}$. Since there are as many moduli as such factors (cf. eq.~\eqref{instanton_degree}) 
we can simply change variables to decouple the integrals:
\be
I_3=\int \prod_{m=1}^k dz_{(m)}^3 \, \prod_{i\in \{V_{-1}\}} \frac{1}{\xi_i{\bar\mu}^3(\z_i)} =\frac{1}{\det_{i\in\{V_{-1}\} }(\xi_i \z_i^{m-1}) }  
\int \prod_{m=1}^k \frac{dz_{(m)}^3}{z_{(m)}^3} 
\ .
\label{mubar3_int}
\ee
In general, the $k\times k$ matrix in the denominator is non-singular. We also notice that for this choice of vertex operators we must choose the
integration contour for $z_{(m)}^3$ to enclose the origin of the ${\bar\mu}^3$ moduli space.
Thus, the integral over the  ${\bar\mu}^3$ moduli is 
\footnote{
We note that the analogous integral appearing if we choose to trade $V_{-1}$ for $V_\delta$ has the same value. Indeed, in this case $I_3$ is replaced with
$$
I'_3=\int \prod_{m=1}^k dz_{(m)}^3 \, \prod_{i\in \{V_\delta\}} \delta(\xi_i{\bar\mu}^3(\z_i)) = \frac{1}{\det_{i\in\{V_{\delta}\} }(\xi_i \z_i^{m-1}) } \ .
$$
}
\be
I_3 = \frac{1}{\det_{i\in\{V_{-1}\} }(\xi_i \z_i^{m-1}) }  \ .
\label{I3}
\ee

\

To evaluate the remaining integrals we first examine the consequences of the special kinematic configuration~\eqref{external_state_condition_av}. 
As discussed in the Introduction and in \S~2, to extract a physically-meaningful amplitude it is necessary to strip off the singular and 
vanishing factors \eqref{expect_singular} and \eqref{vanishing_factor}, respectively. It is not difficult to see that the factor \eqref{expect_singular}
appears quite naturally. Indeed, evaluating the integral over the ${\bar\lambda}$ moduli without imposing this restriction and carrying 
out manipulations similar to those in \cite{Roiban:2004yf} it is possible to extract an overall factor of 
\be
\delta^4(\sum_{i=1}^{2n} {\bar\mu}_i^a{\mu}_{i{\dot a}}) \ ;
\ee
Then, the constraint \eqref{external_state_condition_av} sets to zero the antisymmetric part of the argument of the delta function and thus
it yields \eqref{expect_singular}.  If the constraint \eqref{external_state_condition_av} is imposed strictly before the integration over moduli 
is carried out it yields a divergence proportional to the volume of some integration variable. We will therefore first isolate the 
origin of this singularity, which may be identified as the appearance of a shift symmetry, and account for the Jacobian relating 
it to \eqref{expect_singular}. 

The relevant integral is 
\be
I_\text{exp}=\int \prod_{m=1}^k d^2 z_{(m)}^{\dot a}\prod_{i=1}^{2n} \delta^2(\bar{\mu}_i-\xi_i\bar{\mu}(\z_i))
e^{i\mu_{i\dot{a}}\xi_i\bar{\lambda}^{\dot{a}}(\z_i)} \ .
\ee
In the presence of  eq.~\eqref{external_state_condition_av} the integrand is invariant under the shift transformations
\be
{\bar\lambda}^{\dot 1}(\z)\mapsto {\bar\lambda}^{\dot 1}(\z) + a {\bar\mu}^{1}(\z)
\qquad
{\bar\lambda}^{\dot 2}(\z)\mapsto {\bar\lambda}^{\dot 2}(\z) + a {\bar\mu}^{2}(\z) \ ;
\ee
This symmetry allows one to set to zero one of the integration variables. 

To isolate the problem we trade the integral over one modulus (which we set to zero using the shift transformation) to an integral 
over the collective coordinate $a$. The integral becomes 
\be
I_\text{exp} = 
\int    \left(\prod_{m=1}^k\prod_{{\dot a}=1}^2\right)' d z_{(m)}^{\dot a}\prod_{i=1}^{2n} \delta^2(\bar{\mu}_i-\xi_i\bar{\mu}(\z_i))
e^{i\mu_{i\dot{a}}\xi_i{\hat {\bar{\lambda}}}^{\dot{a}}(\z_i)} \int (z_{(1)}^{1}{d a}) e^{i a \sum_{i=1}^n {\bar\mu}_i^a{\mu}_{i{\dot a}}\delta_a^{\dot a}}
\ee
where ${\hat{\bar{\lambda}}} = {\bar{\lambda}}|_{z_{(1)}^{\dot 1} =0}$ and the prime in the measure of the first integral signals that one is not 
supposed to integrate over $z_{(1)}^{\dot 1}$. The second integral gives the delta function \eqref{expect_singular} which
is responsible for the singularity mentioned above and the Jacobian factor from the change of variables from $z_{(1)}^{\dot 1}$ to $a$.  

In the first integral we can safely impose the projection \eqref{external_state_condition_av}:
\be
\prod_{i=1}^{2n} \delta^2(\bar{\mu}_i-\xi_i\bar{\mu}(\z_i))e^{i\mu_{i\dot{a}}\xi_i{\hat {\bar{\lambda}}}^{\dot{a}}(\z_i)}
\stackrel{\text{eq}~\eqref{external_state_condition_av}}{-\!\!\!-\!\!\!-\!\!\!\longrightarrow} 
\prod_{i=1}^{2n} \delta^2(\bar{\mu}_i-\xi_i\bar{\mu}(\z_i))e^{i\xi_i^2\sum_{m,l=1}^k (z_{(m)1}{\hat z}^{\dot{1}}_{(l)}
+z_{(m)2}{\hat z}^{\dot{2}}_{(l)})\z_i^{m+l-2}} \ ,
\label{useconstraint}
\ee
where ${\hat z}^{\dot{a}}_{(l)}={z}^{\dot{a}}_{(l)}$ unless ${\dot a}=1$ and $l=1$ when it is zero.
As mentioned earlier, the identification \eqref{external_state_condition_av} breaks the manifest 
$\text{SL}(2,\IR)\times \text{SL}(2,\IR)$ of \eqref{BigSym} to a single (diagonal) $\text{SL}(2,\IR)$ which is manifest in the 3d twistor 
space. This breaking is manifest in the equation above; we shall use the shorthand notation  
$(z_{(m)1}{\hat z}^{\dot{1}}_{(l)}+z_{(m)2}{\hat z}^{\dot{2}}_{(l)})\equiv z_{(m)a}{\hat z}^{\dot{a}}_{(l)}\delta^a_{\dot a}$. 

To carry out the integral over the moduli ${\hat z}_{(m)}^{\dot a}$ it is useful to change variables to 
\begin{align}
\alpha_w=\sum_{m=1}^k\sum_{l=1}^k \, \delta^a_{\dot a}z_{(m)a}{\hat z}^{\dot{a}}_{(l)}\delta_{m+l-1,w} \ ;
\end{align}
%
%
the Jacobian of this transformation is
\be
\det \frac{\partial\alpha_w}{\partial {\hat z}^{\dot{a}}_{(l)}}
=\det \sum_{m=1}^kz_{(m)a}\delta_{m+l-1,w}
=\det{}_{w\times (l,a)}(z_{(w-l+1)a}) \ ,
\label{determinant skifte alpha}
\ee
where we indicated the indices of the matrix whose determinant is to be evaluated.
Their range is $w=1,\dots, 2k-1$, $l=1,\dots,k$, $a=1,2$ and the pair $(l, a) = (1, 1)$ is not included. 
The entries of the matrix corresponding to unphysical values of $w-l+1$ are zero.
The exponents in eq.~\eqref{useconstraint} are linear in $\alpha_w$ and thus 
the $\alpha_w$-integrals yield only delta functions; $I_\text{exp}$ becomes
\be
I_\text{exp}=\frac{z_{(1)}^{1}}{\det_{w\times (l,a)}(z_{(w-l+1)a})} \prod_{w=1}^{2k-1}\delta\left(\sum_{i=1}^{2n} \xi_i^2\rho_i^{w-1}\right)
\delta(\sum_{i=1}^{2n} {\bar\mu}_i^a{\mu}_{i{\dot a}}\delta_a^{\dot a})
\prod_{i=1}^{2n} \delta^2(\bar{\mu}_i-\xi_i\bar{\mu}(\z_i)) \ .
\label{Delta1} 
\ee

\

The last component of  ${\cal A}_{2n, k}$ is the integral over the fermionic moduli $z_{(l)}^4$ and $z_{(l)}^5$:
\be
I_{\text{exp}} I_{4,5} = I_\text{exp}\,
\int \prod_{l=1}^kdz_{(l)}^4dz_{(l)}^5 \prod_{i=1}^{2n}\delta(\xi_i \bar{\psi}^4(\rho_i){\mu}_{i \dot 1}+\xi_i\bar{\psi}^5(\rho_i){\mu}_{i \dot 2}) \ .
\label{I45}
\ee
We included $I_\text{exp}$ because the delta functions present in it are relevant for the properties of the integrand. We note that, 
similarly to $I_{\text{exp}}$, in the presence of the kinematic constraint \eqref{external_state_condition_av} this integrand acquires 
the fermionic shift symmetry
\be
 \bar{\psi}^4 \mapsto  \bar{\psi}^4 + \eta  \bar{\mu}^1
 ~,\quad
\bar{\psi}^5 \mapsto  \bar{\psi}^5 + \eta  \bar{\mu}^2 \ .
\ee
Thus, one modulus can be set to zero and consequently $I_{4,5}$ vanishes identically (since it is a Grassmann integral). 
To extract this zero we carry out similar manipulations as for $I_\text{exp}$  and isolate the integration variable that disappears 
should we impose the strict identification \eqref{external_state_condition_av}. First we trade the integral over $z_{(1)}^4$ for an 
integral over $\eta=z_{(1)}^4/z^1_{(1)}$ 
\be
I_{\text{exp}} I_{4,5} = I_\text{exp}\,
\int\frac{d\eta}{z_1^1} \prod_{l=2}^kdz_{(l)}^4\prod_{m=1}^kdz_{(m)}^5 \prod_{i=1}^{2n}
\delta(\xi_i {\hat {\bar{\psi}}}^4(\rho_i){\mu}_{i \dot 1}+\xi_i\bar{\psi}^5(\rho_i){\mu}_{i \dot 2} 
+\eta {\bar\mu}_i^a{\mu}_{i{\dot a}}\delta_a^{\dot a}) \ .
\ee
Carrying out the integration over all fermionic moduli except $\eta$ leads to a determinant
\be
\label{supermomentum determinant nxn 0}
I_{\text{exp}} I_{4,5} = I_{\text{exp}}\,
\det \frac{\partial(\xi_i({\mu}_{i {\dot 1}}\sum_{m=2}^kz_{(m)}^4\z_i^{m-1}+{\mu}_{i{\dot 2}}\sum_{m=1}^kz_{(m)}^5\z_i^{m-1}))_{i=2,\dots,2n}}
{\partial z^{A}_{(m)}|_{A=4,5}}\int \frac{d\eta}{z_{(1)}^1} \delta(\eta\sum_{i=1}^{2n}{\bar\mu}_i^a{\mu}_{i{\dot a}}\delta_a^{\dot a}) \ ,
\nonumber
\ee
while the $\eta$-integral produces a regularized zero and the corresponding Jacobian factor:
\be
I_{\text{exp}} I_{4,5} =I_{\text{exp}}\,\det(\xi_i {\mu}_{i{\dot 1}} \z^{m-1}_i|_{m=2,\dots, k}; \xi_i {\mu}_{i{\dot 2}} \z^{m-1}_i|_{m=1,\dots, k}) \, \frac{1}{z_{(1)}^1}
(\sum_{i=1}^n{\bar\mu}_i^a{\mu}_{i{\dot a}}\delta_a^{\dot a})\ .
\ee
Here $i$ takes $2n-1$ values as one fermionic delta function was treated separately. We have therefore extracted the origin of the zero 
value of the integral in the limit \eqref{external_state_condition_av} and it is parametrized by the same quantity leading to the divergence 
of the naive bosonic moduli integral in the same limit. For the remaining determinant we can enforce \eqref{external_state_condition_av} 
and find
\be
\label{supermomentum determinant nxn}
I_{\text{exp}}I_{4,5} = 
I_{\text{exp}}\,\det(\xi_i^2 \sum_{l=1}^k \z^{m+l-2}_i z_{(l)}^{1} |_{m=2,\dots, k}; \xi_i^2 \sum_{l=1}^k \z^{m+l-2}_i z_{(l)}^{2}|_{m=1,\dots, k}) 
 \; \frac{1}{z_1^1} (\sum_{i=1}^{2n}{\bar\mu}_i^a{\mu}_{i{\dot a}}\delta_a^{\dot a})\ .
\ee
The $z_{(1)}^1$ factor cancels a similar factor in $I_\text{exp}$; moreover the dependence on the $z_{(m)}^a$ moduli cancels between the 
determinant in $I_\text{exp}$ and the one that appeared from the fermionic integration. To see this one simply changes the summation index
to $w=l+m-1$ with the appropriate upper bound on its range; the result of this cancellation is:
\be
\label{extraints}
I_{\text{exp}}I_{4,5} &=&  
\det{}_{w\times i}(\xi_i^2 \z_i^{w-1}) \prod_{w=1}^{2k-1}\delta\left(\sum_{i=1}^{2n}\xi_i^2\rho_i^{w-1}\right)
\prod_{i=1}^{2n} \delta^2(\bar{\mu}_i-\xi_i\bar{\mu}(\z_i))
\\
&&\times (\sum_{i=1}^{2n}{\bar\mu}_i^a{\mu}_{i{\dot a}}\delta_a^{\dot a})\delta( \sum_{i=1}^{2n}{\bar\mu}_i^a{\mu}_{i{\dot a}}\delta_a^{\dot a} )
\nonumber
\ee
where 
$i=2, \dots, 2n$ and $w=1,\dots, 2k-1=1,\dots, 2n-1$.

Putting together eqs.~\eqref{theamplitude}, \eqref{Aparametrization}, \eqref{mubar3_int} and \eqref{extraints} we find that the 
scattering amplitudes of $q_{\text{U}(1)}=0$ states in the string theory constructed in \S~\ref{string_th} are
\be
{\cal A}_{2n, n} 
&=& \int \frac{\prod_{i=1}^nd\z_i}{\Vol\,\text{GL}(2)} \frac{d\xi_i}{\xi_i} \prod_{m=1}^{n} d^{2|3}dz_{(m)}^I
\frac{ \det{}_{w\times i}(\xi_i^2 \z_i^{w-1})}{\det{}_{(i\in\{V_{-1} \}) \times m}(\xi_i \z_i^{m-1})} \prod_{w=1}^{2k-1}\delta\left(\sum_{i=1}^{2n}\xi_i^2\rho_i^{w-1}\right)
\; 
\cr
&&~~\times\prod_{i=1}^{2n} \delta^{2|0}(\bar{\mu}_i-\xi_i\bar{\mu}(\z_i))\delta^{0|3}({\bar\eta}_i^A - \xi_i \bar{\psi}^A(\z_i)) \, \langle J_{12}(\z_1)\dots J_{21}(\z_{2n}) \rangle
\cr
&&~~\times (\sum_{i=1}^{2n}{\bar\mu}_i^a{\mu}_{i{\dot a}}\delta_a^{\dot a})\delta( \sum_{i=1}^{2n}{\bar\mu}_i^a{\mu}_{i{\dot a}}\delta_a^{\dot a} ) \ .
\label{theamplitude_massaged}
\ee
Upon dropping the factors on the third line (or, alternatively, defining the kinematic constraints to also include the integral 
operator $\int d\ln (\sum_{i=1}^{2n}{\bar\mu}_i^a{\mu}_{i{\dot a}}\delta_a^{\dot a})$), focusing on the single-trace 
component of the current correlator, making the change of variables $\xi_i = {\tilde\xi}_i^{k-1}$, slightly reorganizing the
integration variables and Fourier-transforming $({\bar\mu}_i^a,{\bar\eta}_i^A)$ to $\Lambda_i=(\lambda_{ia},\eta_{iA})$, 
eq.~\eqref{theamplitude_massaged} becomes~\footnote{A numerical factor similar to the one on the left-hand side exists 
also in the relation between the different formulations of the tree-level amplitudes of $\NeqFour$ sYM theory.}
\be
\frac{{\cal T}_{2n,k}(\Lambda)}{(k-1)^{2n}} &=& \delta_{k, n}
\int 
\frac{d^{2\times 2n}\sigma}{\vol[{\rm GL}(2)]}\;J\;\Delta\;
 \langle J_{12}(\sigma_1)\dots J_{21}(\sigma_n) \rangle\,{\prod_{m=1}^{k}\delta^{2|3}(C_{mi}[\sigma]\Lambda_i)} \ ,
\nonumber\\[2pt]
C_{mi}[\sigma] &=& a_i^{k-m}b_i^{m-1}
~~,\quad
\sigma_i=(a_i,b_i)=\tilde\xi_i(1,\z_i)
~~,\quad
\langle\Psi_p(\sigma_i){\bar\Psi}_q(\sigma_j)\rangle=\frac{\delta_{pq}}{(i,j)}
\nonumber\\[2pt]
\Delta&=&\prod_{j=1}^{2k-1} \delta(\sum_i a_i^{2k-1-j}b_i^{j-1})
~~,\quad
J = \frac{\text{Num}}{\text{Den}}
~~,\quad
(i,j) = a_i b_j - a_j b_i
\label{Lee_Huang}
\\
\text{Den}&=&\prod_{1\le i<j\le k} (2i-1,2j-1)
~~,\quad
\text{Num}=\det_{1\le i,j\le 2k-1}(a_i^{2k-1-j}b_i^{j-1}) = \prod_{1\le i<j\le 2k-1} (i,j) \ ,
\nonumber
\ee
{\it i.e.} the expression  proposed in~\cite{Huang:2012vt} for the tree-level scattering amplitudes  of the ABJ(M) theory. 
The change of variables $\xi_i = {\tilde\xi}_i^{k-1}$ maps directly the various factors in 
eq.~\eqref{theamplitude_massaged} into $\Delta$, $\text{Num}$  and $\text{Den}$.

In the next section we will explore the double-trace structures included in \eqref{theamplitude_massaged} and interpret them 
as the contribution of intermediate ${\cal N}=6$ conformal supergravity states.


\section{ABJ(M) coupling to conformal supergravity \label{CSG}}

Conformal supergravity (CSG) in three dimensions has no propagating degrees of freedom; as discussed in earlier sections, 
this is mirrored in our construction by the fact that there are no vertex operators that do not carry gauge indices.  Thus, 
in three dimensions, the presence of conformal supergravity is observed in the existence of tree-level multi-trace amplitudes 
of colored fields.  
These amplitudes are given by the multi-trace color structures that appear in the current correlator \eqref{current_correlator} 
entering eq.~\eqref{theamplitude_massaged}. In this section we present evidence that they are the same as the multi-trace 
amplitudes following from the Lagrangian of the ABJ(M) theory coupled to ${\cal N}=6$ conformal supergravity constructed 
in~\cite{Chu:2009gi,Chu:2010fk,Nishimura:2013poa}. 

The relation between single-trace and multi-trace amplitudes follows quite straightforwardly from the structure of current correlator;
in the variables of the eq.~\eqref{theamplitude_massaged}, breaking one trace into two traces at positions $i$ and $j$ 
\be
\Tr[T^{a_1}\dots T^{a_j}\dots T^{a_i}\dots T^{a_{2n}}]\longrightarrow \Tr[T^{a_j}\dots T^{a_i}]\Tr[T^{a_1}\dots T^{a_{j-1}}T^{a_{i+1}}\dots T^{a_{2n}}]
\ee
is reflected at the kinematic level by the 
appearance of the multiplicative factor
\be
F_{ij}=\frac{(\z_i-\z_{i+1})(\z_{j-1}-\z_{j})}{(\z_i-\z_j)(\z_{j-1}-\z_{i-1})} \ ,
\label{us_breaktrace}
\ee
or, in the variables of eq.~\eqref{Lee_Huang},
\be
F_{ij} = \frac{(i, i+1)(j-1, j)}{(ij)(j-1, i+1)} \ .
\label{LH_breaktrace}
\ee
The notation\footnote{This notation does not uniquely specify a general multi-trace amplitude; it will however be sufficient for the purpose of
our discussion.} we will use for multi-trace color-stripped amplitudes is $A_{(n_1,n_2,\dots),k}(1,\dots, 2n)$ where $k$ is the instanton degree
that determines it 
and $n_i$ are the numbers of gauge group generators in each trace.
Let us discuss first the case of the four-point amplitudes; this is the simplest example, in that the entire super-amplitude is determined by a
single component amplitude.  

\subsection{The four-point amplitudes}

The single-trace four-point superamplitude arising from \eqref{Lee_Huang} was computed in \cite{Lee:2010du} and shown to 
agree with the ABJ(M) four-point superamplitude.  Its double-trace version can be easily reconstructed from the details of that 
solution after the factor \eqref{LH_breaktrace} is included. 
We will choose to consider the single- and double-trace structures $\Tr[T^{a_1}T^{a_2}T^{a_3}T^{a_4}]$ and 
$\Tr[T^{a_1}T^{a_2}]\Tr[T^{a_3}T^{a_4}]$ ({\it i.e.} we choose $i=1$ and $j=2$).
Gauge fixing GL$(2)$ as in \cite{Lee:2010du}
\begin{align}
C_{11}=c_{12}~,&& C_{12}=1~,&& C_{13}=c_{32}~,&& C_{14}=0 \ , \\
C_{21}=c_{14}~,&& C_{22}=0~,&& C_{23}=c_{34}~,&& C_{24}=1 \ ,
\end{align}
implies that the relevant multiplicative factor \eqref{LH_breaktrace} is
\be
F_{12} = \frac{(C_{14}C_{21} - C_{24}C_{11})(C_{12}C_{23} - C_{22}C_{13})}
       {(C_{14}C_{23} - C_{24}C_{13})(C_{12}C_{21} - C_{22}C_{11})} = \frac{c_{12} c_{34}}{c_{32} c_{14}} \ .
\ee
Using the solution to the bosonic part of the delta-function constraints,
\be
c_{12} =-\frac{\langle 32\rangle}{\langle 31\rangle}
\quad, \quad
c_{32} =-\frac{\langle 21\rangle}{\langle 31\rangle}
\quad, \quad
c_{14} =-\frac{\langle 34\rangle}{\langle 31\rangle}
\quad, \quad
c_{34} =-\frac{\langle 41\rangle}{\langle 31\rangle} \ ,
\ee
we find that the $\Tr[T^{a_1}T^{a_2}]\Tr[T^{a_3}T^{a_4}]$ superamplitude is
\be
A_{(2,2),2}(1,2;3,4) = -\frac{\langle 23\rangle\langle 14\rangle}{\langle 12\rangle\langle 34\rangle}A_{(4,0),2}(1,2,3,4)=
-\frac{\langle 14\rangle}{\langle 34\rangle^3} \delta^3(\sum_{i=1}^4p_i)\delta^6(\sum_{i=1}^4 \lambda_i^\alpha \eta_i^A) \ .
\label{4pt_2trace}
\ee
The presence of the high power of $\langle 34\rangle$ in the superamplitude above can be understood as being due to the
higher-derivative action of the conformal graviton. For component amplitudes having other intermediate fields the power of 
$\langle 34\rangle$ is lowered by the contributions of the fermionic delta function.

Let us consider the four-scalar double-trace amplitude $A_{(2,2), 2}(p_1^{\phi}, p_2^{{\bar\phi}^{23} },p_3^{\phi_{23}}, p_4^{{\bar\phi}})$  as it is the easiest to 
calculate from the Lagrangian and probes the couplings of the conformal supergravity and ABJ(M) fields. From \eqref{4pt_2trace} we can extract its expression:
\be
A_{(2,2), 2}(p_1^{\phi}, p_2^{{\bar\phi}^{23} },p_3^{\phi_{23}}, p_4^{{\bar\phi}}) =
\int \prod_{i=1}^4 d^3\eta_i\; \eta^1_2 (\eta_3^2 \eta_3^3 ) (\eta^1_4\eta^2_4\eta^3_4) \,A_{(2,2),2}(1,2;3,4)
=- \frac{\langle 23 \rangle\langle 3 1\rangle}{\langle 12 \rangle}  \ .
\ee

Inspecting the Lagrangian in ref.~\cite{Nishimura:2013poa} it is not hard to see that 
this amplitude should be generated by a single Feynman graph with the exchange of the Chern-Simons vector field
$B_\mu^{{\cal I}{\cal J}}$ (the SO$(6)$ R-symmetry gauge field) in
\be
\label{CSGlagrangian}
{\cal L} &=& -e D_\mu {\bar \phi}^i_{\cal A} D^\mu {\phi}_i^{\cal A}+\dots
\\
D_\mu {\phi}_i^{\cal A} &=& (\partial_\mu +\frac{1}{2} i B_\mu)\phi_i^{\cal A} -\frac{1}{4} B_\mu^{{\cal I}{\cal J}}\phi_i^{\cal B} 
(\Sigma^{{\cal I}{\cal J}})_{\cal B}{}^{\cal A} - \phi_j^{\cal A} {\tilde A}_\mu{}^j{}_i \ ,
\nonumber
\ee
where ${\cal I}$ and ${\cal J}$ are SO$(6)$ indices in the fundamental representation, ${\cal A}$ and ${\cal B}$ are SU$(4)$ fundamental 
indices and $(\Sigma^{{\cal I}{\cal J}})_{\cal B}{}^{\cal A}$ are the SO$(6)$ generators in the spinor representation. The relation of the scalar fields 
carrying the scalar fields $SU(4)$ indices \cite{Nishimura:2013poa}  and the fields in the on-shell multiplets~\eqref{superfields} follows from the embedding of
the manifest SU$(3)$ on shell R-symmetry group in SU$(4)$:
\be
(\phi^1, \phi^2, \phi^3, \phi^4)\sim (\phi_{23}, \phi_{31}, \phi_{12}, \phi)
~~,\qquad
(\bar{\phi}_1, \bar{\phi}_2, \bar{\phi}_3, \bar{\phi}_4)\sim (\phi_1, \phi_2, \phi_3, \phi_{123}) \ .
\ee

Using the Chern-Simons propagator and the standard current interaction Feynman rules following from eq.~\eqref{CSGlagrangian} is it 
not difficult to find
\be
A_{(2,2), 2}(p_1^{\phi^4}, p_2^{{\bar\phi}_1 },p_3^{\phi^1}, p_4^{{\bar\phi_4}}) = 
     (p_1-p_2)^a \frac{\epsilon_{abc}(p_1+p_2)^b}{(p_1+p_2)^2} (p_3-p_4)^c =-
     \frac{\langle 23 \rangle\langle 3 1\rangle}{\langle 12 \rangle} \ ,
\ee
matching the result above. Supersymmetry then guarantees that the other components of the superamplitude \eqref{4pt_2trace} follow from the 
same Lagrangian\footnote{It is also easy to check that the four-fermion amplitude 
$\int \prod_{i=1}^4 d^3\eta_i\; (\eta^1_1\eta^2_1\eta^3_1) (\eta^2_1\eta^3_1) \eta_3^1  \,A_{(2|2),2}(1,2;3,4)$ matches 
as well. This amplitude is determined by the minimal coupling 
$$
D_\mu \psi_{{\cal A}i} = (\partial_\mu +\frac{1}{4}\omega_{\mu \, ab}\gamma^{ab} +\frac{1}{2}iB_\mu) \psi_{{\cal A}i}
+\frac{1}{4} B_\mu^{{\cal I}{\cal J}} (\Sigma^{{\cal I}{\cal J}})_{\cal A}{}^{\cal B} \psi_{{\cal B}i}- \psi_{{\cal A} j} {\tilde A}_\mu{}^j{}_i
$$
of ABJ(M) fermions with the same CSG R-symmetry gauge field.}.

\subsection{CGS interactions for any number of external legs}

For a more systematic comparison with the amplitudes following from the Lagrangian of \cite{Nishimura:2013poa} a different approach,
which avoids the direct computation of amplitudes, is more efficient. To this end let us examine again the four-point amplitudes, determined 
by an $k=2$ instanton,  and notice that certain single- and double-trace component amplitudes with {\it different} external states have the same 
expression. For example, 
the following integrals with different  integration measures and integrand denominators are equal:
\begin{align}
\label{equal_integrals}
&\int d\eta^1_1d\eta^1_2d\eta^2_3d\eta^2_4\frac{\prod_{m=1}^2\delta(\sum_{i=1}^4C_{mi}\eta_i^1)\delta(\sum_{i=1}^4C_{mi}\eta_i^2)}
{(12)(21)(34)(43)}
\propto {}\frac{1}{(12)(34)}
\\
&\int d\eta^2_1d\eta^1_2d\eta^1_3d\eta^2_4\frac{\prod_{m=1}^2\delta(\sum_{i=1}^4C_{mi}\eta_i^1)\delta(\sum_{i=1}^4C_{mi}\eta_i^2)}
{(12)(23)(34)(41)}
\propto {}\frac{1}{(12)(34)} \ .
\nonumber
\end{align}
They represent contributions to different double-trace $\Tr[T^{a_1}T^{a_2}]\Tr[T^{a_3}T^{a_4}]$ and single-trace 
$\Tr[T^{a_1}T^{a_2}T^{a_3}T^{a_4}]$ amplitudes, respectively. The denominators included above are the 
only differences between the two superamplitudes; thus, by choosing external states with the same $\eta_1^3, \eta_2^3, \eta^3_3$ 
and $\eta^3_4$ content, the resulting single-trace and double-trace amplitudes are equal. 

{}From a Lagrangian perspective this equality is a consequence of the existence of similar terms  with and without conformal supergravity 
interactions,
as detailed below:
\be
\begin{array}{|c|c|}
\hline
\multicolumn{2}{|c|}{\int d\eta_1^3 d\eta_2^3 d\eta_3^3 d\eta_4^3\;\eta_3^3\eta^3_4 \vphantom{\Big|} } \\
\hline
\mathrm{Tr}\left(\bar{\phi}_1(1)\psi_1(2)\right)\mathrm{Tr}\left(\bar{\psi}^1(3)\phi^1(4)\right) \vphantom{\Big|} &
\mathrm{Tr}\left(\bar{\phi}_2(1)\psi_1(2)\bar{\psi}^2(3)\phi^1(4)\right)\\
\hline
\multicolumn{2}{|c|}{\textrm{Contact term} \vphantom{\Big|}}\\
\hline
-\frac{3}{8}ge\bar{\psi}^{Ai}\psi_{Bj}\bar{\phi}^j_A\phi^B_i-\frac{1}{8}ge\bar{\psi}^{Ai}\psi_{Aj}\left(\bar{\phi}\phi\right)^j_i  \vphantom{\Big|}&
-2ef^{ij}_{\phantom{ij}kl}\bar{\psi}^{Ak}\psi_{Bi}\bar{\phi}^l_A\phi^B_j\\
\hline
\multicolumn{2}{|c|}{\int d\eta_1^3 d\eta_2^3 d\eta_3^3 d\eta_4^3\; \eta_2^3\eta^3_4 \vphantom{\Big|}}\\
\hline
\mathrm{Tr}\left(\bar{\phi}_1(1)\phi^2(2)\right)\mathrm{Tr}\left(\bar{\phi}_2(3)\phi^1(4)\right)    \vphantom{\Big|}  &
\mathrm{Tr}\left(\bar{\phi}_2(1)\phi^2(2)\bar{\phi}_1(3)\phi^1(4)\right)\\
\hline
\multicolumn{2}{|c|}{\textrm{Exchange of spin-1 Chern-Simons field with $q=p_1+p_2$ \vphantom{\Big|}}}\\
\hline
-e{D}_\mu\bar{\phi}^i_A{D}^\mu \phi^A_i&-e{D}_\mu\bar{\phi}^i_A{D}^\mu \phi^A_i  \vphantom{^{\big|}}\\
{D}^\mu \phi^A_i=\cdots-\frac{1}{4}B^{IJ}_\mu \phi^B_i(\Sigma)^A_{\phantom{A}B}&{D}^\mu \phi^A_i=\cdots-\phi^A_j\tilde{A}^{\phantom{\mu}j}_{\mu\phantom{j}i}  \vphantom{\Big|}\\
\hline
\multicolumn{2}{|c|}{\int d\eta_1^3 d\eta_2^3 d\eta_3^3 d\eta_4^3\; \eta_1^3\eta^3_3 \vphantom{\Big|}}\\
\hline
\mathrm{Tr}\left(\bar{\psi}^2(1)\psi_1(2)\right)\mathrm{Tr}\left(\bar{\psi}^1(3)\psi_2(4)\right) \vphantom{\Big|} &
\mathrm{Tr}\left(\bar{\psi}^1(1)\psi_1(2)\bar{\psi}^2(3)\psi_2(4)\right)\\
\hline
\multicolumn{2}{|c|}{\textrm{Exchange of spin-1 Chern-Simons field with $q=p_1+p_2 \vphantom{\Big|}$ }}\\
\hline
-\frac{e}{2}\left(\bar{\psi}^{Ai}\gamma^\mu{D}_\mu\psi_{Ai}-{D}_\mu\bar{\psi}^{Ai}\gamma^\mu\psi_{Ai}\right)&-\frac{e}{2}\left(\bar{\psi}^{Ai}\gamma^\mu{D}_\mu\psi_{Ai}-{D}_\mu\bar{\psi}^{Ai}\gamma^\mu\psi_{Ai}\right)  \vphantom{^{\big|}}\\
{D}_\mu\psi_{Ai}=\cdots+\frac{1}{4}B^{IJ}_\mu(\Sigma^{IJ})_A^{\phantom{A}B}\psi_{Bi}&{D}_\mu\psi_{Ai}=\cdots-\psi_{Aj}\tilde{A}_{\mu\phantom{j}i}^{\phantom{\mu}j}  \vphantom{{\Big|}}\\
\hline
%
\multicolumn{2}{|c|}{\int d\eta_1^3 d\eta_2^3 d\eta_3^3 d\eta_4^3\; \eta_1^3\eta^3_4 \vphantom{\Big|}}\\
\hline
\mathrm{Tr}\left(\bar{\psi}^2(1)\psi_1(2)\right)\mathrm{Tr}\left(\bar{\phi}_2(3)\phi^1(4)\right)   \vphantom{\Big|}&
\mathrm{Tr}\left(\bar{\psi}^1(1)\psi_1(2)\bar{\phi}_1(3)\phi^1(4)\right)\\
\hline
\multicolumn{2}{|c|}{\textrm{Contact term and exchange of spin-1 Chern-Simons field with $q=p_1+p_2$ \vphantom{\Big|}}}\\
\hline
-e{D}_\mu\bar{\phi}^i_A{D}^\mu \phi^A_i&-e{D}_\mu\bar{\phi}^i_A{D}^\mu \phi^A_i   \vphantom{^{\big|}}\\
{D}^\mu \phi^A_i=\cdots-\frac{1}{4}B^{IJ}_\mu \phi^B_i(\Sigma)^A_{\phantom{A}B}&{D}^\mu \phi^A_i=\cdots-\phi^A_j\tilde{A}^{\phantom{\mu}j}_{\mu\phantom{j}i}   \vphantom{{\Big|}} \\
-\frac{e}{2}\left(\bar{\psi}^{Ai}\gamma^\mu{D}_\mu\psi_{Ai}-{D}_\mu\bar{\psi}^{Ai}\gamma^\mu\psi_{Ai}\right)&-\frac{e}{2}\left(\bar{\psi}^{Ai}\gamma^\mu{D}_\mu\psi_{Ai}-{D}_\mu\bar{\psi}^{Ai}\gamma^\mu\psi_{Ai}\right)  \\
{D}_\mu\psi_{Ai}=\cdots+\frac{1}{4}B^{IJ}_\mu(\Sigma^{IJ})_A^{\phantom{A}B}\psi_{Bi}&{D}_\mu\psi_{Ai}=\cdots-\psi_{Aj}\tilde{A}_{\mu\phantom{j}i}^{\phantom{\mu}j} \vphantom{{\Big|}} \\
\frac{1}{4}ge\bar{\psi}^{Ai}\psi_{Bi}(\bar{\phi}\phi)_A^{\phantom{A}B}&-2ef^{ij}_{\phantom{ij}kl}\bar{\psi}^{Ak}\psi_{Bi}(\bar{\phi}_A^l\phi_j^B-\frac{1}{2}\delta^B_A\bar{\phi}_C^l\phi_j^C) \vphantom{{\Big|}}\\
\hline
\end{array}
\ee
This comparison probes only the matter interactions of conformal supergravity fields; while it does not probe the self-interactions of the latter, they 
should be uniquely determined by superconformal symmetry.

The discussion above can be extended to a comparison of single- and double-trace amplitudes with any number of external legs. The one
potential difficulty is that, since the instanton degree is larger than $k=2$, it is not obvious that simply interchanging two 
Grassmann coordinates has the same effect as in eq.~\eqref{equal_integrals}; a judicious choice of external states addresses this issue.
As in the beginning of this section, let $i$ and $j$ be the states that are non-adjacent in the single-trace but are adjacent in the double-trace  
amplitudes. 
In the double-trace case we choose $i$ and $j$ to contain the Grassmann variables $\eta_i^1$ and $\eta_j^1$ and  the
states $i+1$ and $j-1$ to contain $\eta_{i+1}^2$ and $\eta_{j-1}^2$ and any  other state $k$ either contains the product $\eta_k^1\eta_k^2$ 
or is independent of $\eta^1$ and $\eta^2$. 
In the single-trace case we interchange the $1$ and $2$ indices of $\eta_j$ and $\eta_{i+1}$ while leaving unchanged all the other dependence 
on $\eta^1$ and $\eta^2$.
This choice of $\eta$-s implies that the fields  $(i, i+1, j-1, j)$ are either $(\bar{\psi}^2, {\psi}_2, \bar{\psi}^1, {\psi}_1)$ or 
$(\bar{\phi}_1, {\phi}^1, \bar{\phi}_2, {\phi}^2)$ in the double-trace amplitude while in the single-trace amplitude 
the fields  $(i, i+1, j-1, j)$ are either $(\bar{\psi}^2, {\psi}_1, \bar{\psi}^1, {\psi}_2)$ or $(\bar{\phi}_1, {\phi}^2, \bar{\phi}_2, {\phi}^1)$.
By explicitly evaluating the integrals analogous to \eqref{equal_integrals} it is not difficult to see that they are equal.

To figure out the gauge-singlet fields that can be responsible for these amplitudes we follow the SU$(4)$ indices. Since the two traces transform nontrivially 
under SU$(4)$ the fields exchanged between them must carry such indices; this excludes several fields, including the (conformal) graviton. 
The (conformal) gravitino as well as some auxiliary fields interact through vertices which are antisymmetric in their SU$(4)$ indices; this interaction
is forbidden because of our assumption that both particles $i$ and $j$ appear together with $\eta^1$ in their respective multiplets. 
Thus, apart from contact terms in the ABJ(M)-coupled conformal supergravity Lagrangian,  the only fields 
that can yield a double-trace structure for the field configuration described above are the \mbox{spin-1} Chern-Simons fields $B^{IJ}_{\mu}$.
Inspecting the field configuration in the single-trace amplitude it is not difficult to conclude that the interaction between the fields which 
belonged to the two different traces can be mediated either by a contact term or by an SU$(4)$-neutral field; the only such option is the 
spin-1 Chern Simons field $A_\mu$. 
These interactions are the same as those leading to the correspondence of four-point single- and double-trace amplitudes, as expected 
from the equality of amplitudes following from our construction.


\section{Outlook \label{outlook}}

In this paper we have shown that a particular truncation of an open string theory on CP$^{2,2|4,1}$ reproduces the tree-level amplitudes 
of the ABJ(M) theory in the form proposed in \cite{Huang:2012vt}. 
%
The spectrum of this string theory contains a finite number of  states and the truncation can be interpreted as dimensional reduction; 
thus, our construction suggests  that the ABJ(M) theory can be interpreted as the dimensional reduction of a higher-dimensional 
(perhaps non-unitary) field theory. 
It would be interesting to find a Lagrangian interpretation of the latter; the S matrix following from the vertex operators \eqref{BigSym} implies that 
it should conserve the six-component matrix ${\bar\mu}^\alpha\mu_{\dot a}$. 
Since CP$^{2,2|4,1}$ is a Calabi-Yau space, this theory can also be described as the string field theory of the topological B-model on this space.

The string theory also contains states carrying no gauge group indices. Upon truncation to a three-dimensional theory with the ABJ(M) field content 
such asymptotic states disappear from the spectrum. They survive, however, from an off-shell perspective and affect the S matrix by generating 
multi-trace scattering amplitudes at tree level. By computing double-trace amplitudes in this string theory we have shown that their interactions 
with the gauge-nonsinglet fields is given by the Lagrangian of the ABJ(M) theory coupled to ${\cal N}=6$ conformal supergravity proposed 
in~\cite{Nishimura:2013poa}.

An important step in our construction, which led to a spectrum containing only the multiplets~\eqref{superfields}, was the truncation to the 
zero-charge states with respect to the current $J_{\text{U}(1)}$ defined in eq.~\eqref{U1_current}. While, as discussed in \S~\ref{string_th}, 
this truncation is consistent at the classical level, quantum mechanically it requires that this current be gauged. To this end there should be 
no second order pole in the $J_{\text{U}(1)}(\z_1)J_{\text{U}(1)}(\z_2)$ OPE, such that inclusion of $J_{\text{U}(1)}$ in the BRST operator 
with the corresponding ghost field $w$ keeps $Q$ nilpotent. Using eq.~\eqref{qBqF}, absence of such a pole requires that
\be
0=q_B^2 - 2q_F^2-\frac{N+M}{NM}q_{}^2 = 2q_F^2 - \frac{N+M}{NM} q_{}^2 \ .
\ee
Thus, with the choice \eqref{fermion_charge} for the $\Psi_{1,2}$ charge, the gauge anomaly cancels only for 
\be
2(N+M) = NM \quad\rightarrow\quad (N,M) \in \{(4,4), (6,3)\}  . 
\ee
It may be possible to accommodate such values of $N$ and $M$ 
while also cancelling the central charge $c=c_{bc}+c_{uv}+c_{ws} = -30$. 

Gauging this symmetry requires that the fermion action be covariantized with respect to the new gauge field. As with the GL$(1)$ gauge field, it 
may be gauged away except for topologically-nontrivial sectors. Thus, the correlation functions of the fermion currents $J_{ij}$ \eqref{BigSym} 
breaks up into a sum over instanton sectors, with the zero-instanton sector  reproducing the amplitudes~\eqref{theamplitude_massaged}. 
Nontrivial instantons will modify the fermion two-point function and thus the current correlators. It would be interesting to understand if such
additional terms can be given an interpretation from a field theory perspective. Such a framework, in which both the kinematic and color part of an 
amplitude are treated symmetrically (and are given by instanton sums), is reminiscent of color/kinematics duality~\cite{Bern:2008qj}. 
The kinematic part however exhibits localization on instantons of fixed degree; since a similar localization does seem to occur for the 
color-fermions, this may be a possible explanation for the absence of the duality for ABJ(M) amplitudes~\cite{Huang:2013kca}. It would be interesting 
to explore whether the duality is restored if one restricts  the "color instantons" to have the same degree as the "kinematic instantons".

A feature of the construction discussed in the previous sections is that, in all trace structures of eq.~\eqref{theamplitude_massaged}, all 
vertex operators of type $V_0$ or $V_{-1}$ are inserted at alternating points. This was, of course, a consequence of their $J_{ij}$ content and 
mirrors the fact that in the ${\cal N}=6$ super-Chern-Simons theories no two adjacent external states in a color-stripped amplitude  
belong to the same multiplet.
For product gauge groups with specific ratios of the rank of the factors it is possible to change this pattern while preserving the truncation 
to $q_{\text{U}(1)}=0$ states at the expense of introducing further fermion fields. For example, with four fermions $(\Psi_1, \Psi_2, \Psi_3, \Psi_4)$ 
with multiplicities $(5N, N, N, N)$ and charges $q_{} = (-1/2, +1/2, +3/2, +1/2)$\footnote{These charges and multiplicities ensure that there is no 
anomalous term in the OPE of the analog of the current $J_F$ and the stress tensor.}, the currents $(J_{21}, J_{12}, J_{14}, J_{43}, J_{32})$ 
have $q_F$ charges $q_F=(-1, +1, +1, +1, -1)$. Defining the operators $V_{0,ij}={\cal P}_FU_0 J_{ij}$ and $V_{-1,ij}={\cal P}_FU_{-1} J_{ij}$, 
the correlators
\be
\langle V_{0,21}(\z_1)V_{-1,12}(\z_2)V_{0,21}(\z_3)V_{-1,12}(\z_4)\rangle 
\quad\text{and}\quad
\langle V_{0,21}(\z_1)V_{-1,14}(\z_2)V_{-1,43}(\z_3)V_{0,32}(\z_4)\rangle 
\label{exotic}
\ee 
are nonvanishing. 
The former leads to the four-point version of eq.~\eqref{theamplitude_massaged}. In the latter the current correlator has a single trace structure which
requires that two multiplets of the same type are adjacent. Moreover, the correlator
\be
\langle V_{0,21}(\z_1)V_{-1,12}(\z_2)V_{0,21}(\z_3)V_{-1,14}(\z_4)V_{-1,43}(\z_5)V_{0,32}(\z_6)\rangle 
\ee
appears to also be nonvanishing and has the correct color structure for the corresponding amplitude to factorize into the product of amplitudes related 
to eq.~\eqref{exotic}.
This example appears to describe a product gauge group with one SU$(5N)$ and three SU$(N)$ factors and matter in bifundamental representation. 
It would be interesting to explore the factorization properties of these and higher-point amplitudes (as well as in more general 
constructions of a similar type), identify the states they factorize on and understand whether there exists a standard quantum field 
theory with this tree-level S-matrix. 

Similarly, by including fermions of different charges it is possible to include states created by a finite number of vertex operators containing factors 
of ${\cal P}_FU_k$ with $k\ne 0,-1$ while continuing to truncate the spectrum to states with vanishing $q_{\text{U}(1)}$ charge\footnote{One may also 
relax this constraint; in this case consistency requires that all states are kept and perhaps the theory is better thought of from a higher-dimensional perspective.}. For example,  
two pairs of fermions with charges $\pm q_1$ and $\pm q_2$ can be used to construct currents with charges $\pm 2q_1$, $\pm 2q_2$, $\pm (q_1+q_2)$
and $\pm (q_1-q_2)$. For suitable choices the relation between $q_{1,2}$ and $q_F$ it may be possible to construct a theory with four gauge groups
(of unrelated ranks) and eight matter multiplets in the bi-fundamental representation. The tree-level scattering amplitudes of such theories can be constructed 
following the analysis in this paper; the high degree poles in the $U_{n}$ factors with $n\le -2$ as well as the potential zeroes in the $U_{n}$ factors with $n\ge 1$
suggest that such amplitudes are very constrained. 

In both constructions above it appears that the S-matrix -- and therefore the field theory generating it -- may have ${\cal N}=6$ superconformal  
symmetry; if so, it would be interesting to understand how it evades the arguments of \cite{Gaiotto:2008sd, Hosomichi:2008jb, Schnabl:2008wj}.

\

\section*{Acknowledgments }

We would like to thank Y. -t~ Huang, H.~Johansson and A.~Tseytlin for useful discussions and Y. -t~ Huang, N. Obers and A.~Tseytlin for 
comments on the draft.  OTE would like to thank the Niels Bohr International Academy for hospitality while this paper was being 
written up.
This work is supported by the US Department of Energy under contract DE-SC0008745. 

\

\end{document}